\documentclass{aa}

\usepackage{graphics}

\begin{document}

   \thesaurus{11(11.05.2; 11.06.2; 11.16.1; 11.19.3; 1.03.3; 13.20.1)}

   \title{Dust emission from the lensed Lyman break galaxy \object{cB58}}
%   \subtitle{}

   \authorrunning{Baker et al.}
%   \titlerunning{Dust emission from cB58}

   \author{A.~J. Baker
    \and D.~Lutz
    \and R.~Genzel
    \and L.~J. Tacconi
    \and M.~D. Lehnert}

   \offprints{A.~J. Baker}
   \mail{ajb@mpe.mpg.de}

   \institute{Max-Planck-Institut f\"ur extraterrestrische Physik,
             Postfach 1312, 85741 Garching, Germany
             }

   \date{Received 4 October 2000 / Accepted 20 April 2001}

   \maketitle

   \begin{abstract}

We detect 1.2\,mm continuum emission from dust in the gravitationally lensed 
Lyman break galaxy \object{MS\,1512+36-cB58}.  Our detected flux is 
surprisingly low: relative to local starburst galaxies, \object{cB58} appears 
to produce somewhat less far-IR emission than its UV reddening predicts.  
After comparing several different estimates of the source's dust content, we 
conclude that the apparent discrepancy is most likely related to uncertainty 
in its UV spectral slope.  Alternate scenarios to account for a far-IR 
``deficit'' which rely on a high dust temperature or differential 
magnification are less satisfactory.  Our result underscores one of the risks 
inherent in characterizing the cosmic star formation history from rest-UV 
data alone.

               
   \end{abstract}

%
%________________________________________________________________

\section{Introduction}

Large samples of actively star-forming galaxies at $z \geq 2$ can now be 
identified through color selection techniques which exploit the Lyman break 
(e.g., Steidel et al. \cite{stei99}).  In many respects, these sources 
resemble UV-bright starburst galaxies at $z = 0$: the two populations have 
similar bolometric surface brightnesses (Meurer et al. \cite{meur97}), 
UV-through-optical spectral energy distributions (SEDs: e.g., Ellingson et al. 
\cite{elli96}; Sawicki \& Yee \cite{sawi98}), and blueshifted interstellar 
absorption from outflowing gas 
(e.g., Pettini et al. \cite{pett98}).  The analogy between Lyman break 
galaxies (LBGs) and local starbursts is not exact, however: LBGs appear to be 
forming stars at rates which are ``scaled up'' by their larger physical sizes 
(Meurer et al. \cite{meur97}).  At longer wavelengths, moreover, the analogy 
can barely be evaluated, since few direct far-IR and submillimeter 
observations of LBGs exist.  Two recent attempts have been made to fill this 
gap.  Chapman et al. (\cite{chap00}) detect only one of sixteen LBGs at 
$450\,\mathrm{\mu m}$ and $850\,\mathrm{\mu m}$ with the Submillimeter 
Common-User Bolometer Array (SCUBA).  Peacock et al. (\cite{peac00}) compare 
\emph{HST} and SCUBA observations of the Hubble Deep Field and achieve a 
statistical detection at $850\,\mathrm{\mu m}$ of the \emph{HST} sources with 
photometric redshifts, but can make no statement about specific objects with 
spectroscopic redshifts.  The limitations of these studies motivated us to 
observe the gravitationally lensed system \object{MS\,1512+36-cB58} (hereafter 
\object{cB58}: Yee et al. \cite{yee96}). Thanks to magnification by the $z = 
0.37$ cluster \object{MS\,1512+36}, this $z = 2.7$ LBG was already 
well-studied in the rest UV and optical (e.g., Ellingson et al. \cite{elli96}; 
Pettini et al. \cite{pett00}; Teplitz et al. \cite{tepl00}), and promised to 
be unusually detectable at longer wavelengths.

\section{Observations}

From 3--7 March 2000, we observed \object{cB58} in on-off photometric mode 
with the 37-element Max-Planck Millimeter Bolometer (MAMBO) array (Kreysa et 
al. \cite{krey98}) at the IRAM 30m telescope.  We pointed the array's central 
pixel at the optical position (J2000 coordinates 15:14:22.20 and +36:36:24.4) 
determined by Abraham et al. (\cite{abra98}).  We chopped the secondary mirror 
by $53^{\prime\prime}$ in azimuth at 0.5\,Hz and nodded by the same 
$53^{\prime\prime}$ throw every ten seconds; the HPBW of our beam was 
$11^{\prime\prime}$ at 1.2\,mm.  We reduced the data with the MOPSI package, 
using all pixels in the first ring around the central pixel for estimating and 
removing correlated fluctuations due to the sky background.  Our 
flux scale of $6250\,\mathrm{counts\,Jy^{-1}}$ was based on observations of 
planets.  If we include all data from our total on+off integration time of 200 
minutes, we achieve a $6.0\sigma$ detection of $1.32 \pm 0.22\,\mathrm{mJy}$.  
For the purposes of this paper, however, we will exclude 60 minutes of data 
from our last observing night, for which pointing and sky background 
instabilities coupled with a large opacity correction ($\tau \sim 0.3$ vs. 
$\leq 0.16$ over the rest of our run) reduce our confidence in the 
photometry.  Including a 10\% uncertainty in our flux scale then yields a 
final [$4.4\sigma$] detection of $1.06 \pm 0.35\,\mathrm{mJy}$.  

\section{A comparison with local starbursts}

\begin{figure}
 \resizebox{\hsize}{!}{\includegraphics{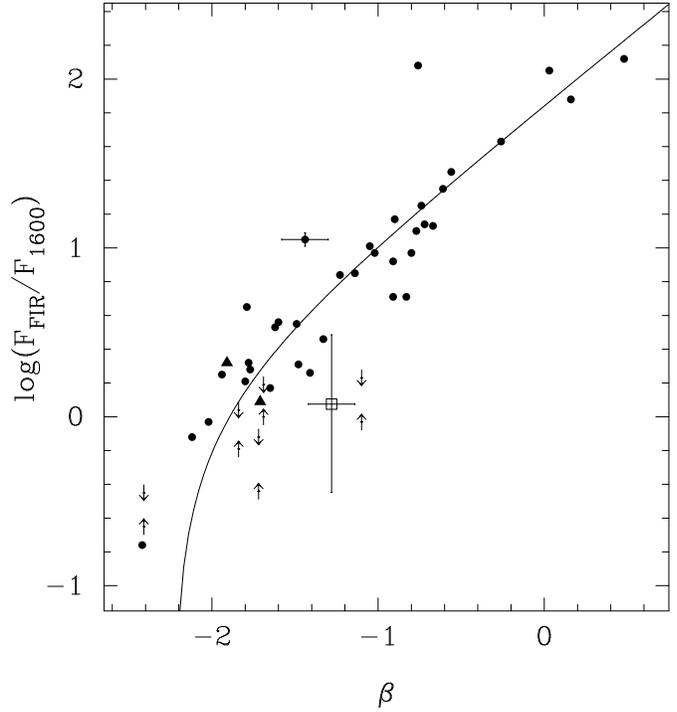}}
 \caption{Far-IR-to-UV flux ratio vs. UV spectral slope.  Estimates for local 
starbursts are triangles (for \emph{IRAS} PSC flux densities), circles (for 
\emph{IRAS} FSC flux densities), or pairs of arrows (for FSC $60\,\mathrm{\mu 
m}$ detections and $100\,\mathrm{\mu m}$ upper limits).  The fit and (for 
\object{NGC\,2537}) representative error bars are taken from Meurer et al. 
(\cite{meur99}).  \object{cB58} is an open square.}
 \label{fig:meur}
\end{figure}

Heckman et al. (\cite{heck98}) and Meurer et al. (\cite{meur99}) have showed 
that local starburst and star-forming galaxies obey scaling relations between 
various global properties and the parameters of their UV spectra.  We can 
therefore test the proposition that \object{cB58} is a ``scaled-up'' local 
starburst by seeing if it fits those relations which are independent of 
distance (and lensing magnification).  A key parameter here is the UV spectral 
slope $\beta$ ($f_\lambda \propto \lambda^\beta$), whose observed value 
indicates the amount of reddening by dust (i.e., from some intrinsically blue  
$\beta_0$) for an assumed star formation history.  Because the same dust which 
reddens stellar UV continua will re-radiate the energy which it absorbs, it is 
natural that UV-bright local systems should obey a relationship between 
$\beta$ and the ratio of far-IR to UV emission.  Meurer et al. (\cite{meur99}) 
determine this relation for a sample which includes only galaxies whose 
starbursts are small enough to fit entirely within the \emph{IUE} aperture.  
They quantify the fraction of reprocessed UV emission as a ratio between
\begin{equation}
F_\mathrm{FIR} = 1.26 \times 10^{-11}[2.58\,f_\nu (60) + f_\nu 
(100)]\,\mathrm{erg\,cm^{-2}\,s^{-1}}
\end{equation}
for $60\,\mathrm{\mu m}$ and $100\,\mathrm{\mu m}$ \emph{IRAS} flux 
densities in Jy, and $F_{1600} = \nu f_\nu$ evaluated at $\nu = 
c/1600\,\mathrm{\AA}$.  Figure~\ref{fig:meur} plots this ratio versus $\beta$ 
for the 43 sources which (a) appear in their Figure 1, and (b) have flux 
densities in the Version 2.0 \emph{IRAS} Faint Source and/or Point Source 
Catalogs.  The local starbursts occupy a predictable locus in the diagram, 
extending from blue, metal-poor systems at the lower left to reddened, dusty 
systems at the upper right.

To place \object{cB58} on this diagram, we first measure $\beta$ from 
the rest-UV spectrum of Pettini et al. (\cite{pett00}), according to the 
prescription of Meurer et al. (\cite{meur99}) for the sake of consistency.  
A least-squares fit (with iterative $\pm 3\sigma$ rejection) to eight spectral 
windows between 1100 and $1975\,\mathrm{\AA}$ gives $\beta = -1.12 \pm 0.10$, 
slightly bluer than the -0.8 ($\beta = -2 -\alpha$ for $f_\nu \propto 
\nu^\alpha$) measured by Pettini et al. (\cite{pett00}) from their fit to 
continuum flux densities at 1300 and $1800\,\mathrm{\AA}$.  Meurer et al. 
(\cite{meur99}) also add an offset $\Delta \beta = -0.16 \pm 0.04$ for each 
source lacking spectral coverage longwards of $1975\,\mathrm{\AA}$; making 
the same adjustment for \object{cB58} yields our final value of $\beta = -1.28 
\pm 0.14$.  Next, we can estimate $F_{1600}$ from the $V_\mathrm{AB} = 20.64 
\pm 0.12$ observed by Ellingson et al. (\cite{elli96}), corresponding to a 
rest wavelength of $1474.9\,\mathrm{\AA}$ (for $z = 2.7290$: 
Teplitz et al. \cite{tepl00}):
\begin{eqnarray}
F_{1474.9} & = & \nu f_\nu(\mathrm{observed}\,5500\,\mathrm{\AA}) \nonumber \\
         & = & 1.10 \pm 0.13 \times 10^{-13}\,\mathrm{erg\,cm^{-2}\,s^{-1}} 
\nonumber
\end{eqnarray}
Scaling this flux by a factor $(1600/1474.9)^{1 + \beta} = 0.98 \pm 0.01$ 
then gives $F_{1600} = 1.07 \pm 0.14 \times 10^{-13}\,\mathrm{erg\,cm^{-2}\,
s^{-1}}$.  

Our final step is to derive $F_\mathrm{FIR}$ from our observed 1.2\,mm 
flux density.  We must multiply the observed $\nu f_\nu = 2.65\times 
10^{-15}\,\mathrm{erg\,cm^{-2}\,s^{-1}}$ by a rest-frame ratio of the form
\begin{equation}
{\frac {F_\mathrm{FIR}}{\nu f_{\nu}}} = {\frac {1.26 \times 10^{-11}}{\nu}}\,
\Big[2.58\,{\frac {f_\nu (60)}{f_\nu (321.8)}} + {\frac {f_\nu (100)}{f_\nu 
(321.8)}}\Big]
\end{equation}
where now $\nu = c\,(1+z)/1.2\,\mathrm{mm} = c/321.8\,\mathrm{\mu m}$.  We must
evaluate the term in brackets by assuming some rest-frame SED for 
\object{cB58}.  Since comparably deep SCUBA observations of \object{cB58} at 
$850\,\mathrm{\mu m}$ have yielded both a detection of $4.2 \pm 
0.9\,\mathrm{mJy}$ (van~der~Werf et al. \cite{vand01}) and a $3\sigma$ upper 
limit of $3.9\,\mathrm{mJy}$ (Sawicki \cite{sawi01}), the 
prospects for an observationally-derived SED are not encouraging.  Instead, we 
simply extend the scaled-up starburst hypothesis for LBGs: we assign to 
\object{cB58} a dust emissivity index $\epsilon = 2$ and dust temperature 
$T_\mathrm{d} = 33.0 \pm 4.5\,\mathrm{K}$ (the mean and dispersion of the 
$60/100\,\mathrm{\mu m}$ color temperatures for the local galaxies in 
Figure~\ref{fig:meur}, as determined from $\epsilon = 2$ modified Planck 
function fits).  We then predict an observed flux in the rest-FIR window 
of $1.28^{+1.63}_{-0.84}\times 10^{-13}\,\mathrm{erg\,cm^{-2}\,s^{-1}}$, with 
large and asymmetric error bars dominated by the uncertainty in $T_\mathrm{d}$.
We emphasize that our extrapolated FIR flux is conservatively large: adoption 
of $\epsilon = 1$ and the corresponding mean $T_\mathrm{d} = 40.7 \pm 
7.1\,\mathrm{K}$ would have lowered our estimate by a factor of two.  We have 
also ignored the potential impact of optically thick dust emission, or of a 
dust component colder than 33.0\,K which could dominate the 
$321.8\,\mathrm{\mu m}$ emission but contribute little to $F_\mathrm{FIR}$.

When plotted in Figure~\ref{fig:meur}, \object{cB58} falls below and/or 
to the right of all objects except \object{UGC\,6456}.  While the scatter in 
the local sample is large, it is strictly true that relative to the fit by 
Meurer et al. (\cite{meur99}), \object{cB58} appears to have too red a UV 
spectral slope for its $F_\mathrm{FIR}/F_{1600}$, and too low a far-IR-to-UV 
ratio (by a factor of $\sim 4.7$) for its $\beta$.  

\section{Four estimates of the dust content in \object{cB58}}

In physical terms, Figure~\ref{fig:meur} suggests that \object{cB58} has 
somewhat more dust than its far-IR emission predicts or somewhat less dust 
than its UV reddening predicts.  We can explore these two possibilities by 
making four distinct estimates of the source's dust content on the basis of 
the local starburst analogy.  We rely in particular on the effective 
attenuation curve $k^\prime(\lambda)$ and other empirical relations presented 
by Calzetti (\cite{calz00}) and references therein.

\vspace{1.0mm}
\noindent
%\begin{underline}
{\bf I. The far-IR-to-UV ratio}
%\end{underline}
%\subsection {The far-IR-to-UV ratio}

From their local starburst sample, Meurer et al. (\cite{meur99}) derive an 
empirical relation between $A(1600\,\mathrm{\AA})$-- the extinction at 
$1600\,\mathrm{\AA}$-- and the $F_\mathrm{FIR}/F_{1600}$ ratio:
\begin{equation}
\mathrm{log}\,(F_\mathrm{FIR}/F_{1600}) = \mathrm{log}\Big(10^{0.4\,
A(1600\,\mathrm{\AA})} - 1\Big) + 0.076
\end{equation}
From our detection of \object{cB58}, we deduce $A(1600\,\mathrm{\AA}) = 
0.75^{+0.63}_{-0.47}$; we then use the Calzetti attenuation curve to evaluate 
the color excess of the stellar continuum, $E_s(B-V) = 
A(\lambda)/k^\prime(\lambda) = 0.075^{+0.063}_{-0.047}$.

\vspace{1.0mm}
\noindent
%\begin{underline}
{\bf II. The H$\mathrm{\alpha}$-to-H$\mathrm{\beta}$ ratio}
%\end{underline}
%\subsection {The $H\alpha$-to-$H\beta$ ratio}

Because ionizing stars are generally younger and more embedded than those 
which dominate the UV continuum, the color excess measured from observations 
of (gas) recombination lines, $E_g(B-V)$, tends to exceed $E_s(B-V)$ and 
can be straightforwardly measured using any optical extinction curve 
$k(\lambda)$.  In \object{cB58}, Teplitz et al. (\cite{tepl00}) report a 
Balmer decrement of $3.23 \pm 0.4$, which for the LMC extinction curve of 
Howarth (\cite{howa86}) implies $E_g(B-V) = 0.12 \pm 0.12$ (rather than the 
former authors' $\sim 0.27$).  Since Calzetti (\cite{calz00}) finds that local 
starbursts obey the proportionality $E_s(B-V) = 0.44\,E_g(B-V)$, we conclude 
$E_s(B-V) \simeq 0.053^{+0.051}_{-0.053}$.

\vspace{1.0mm}
\noindent
%\begin{underline}
{\bf III. The UV line spectrum and UV-to-H$\mathrm{\alpha}$ ratio}
%\end{underline}
%\subsection{The UV line spectrum and UV-to-$H\alpha$ ratio}

The rest-UV spectrum of \object{cB58} obtained by Pettini et al. 
(\cite{pett00}) is of a quality which permits detailed population synthesis 
modelling.  Pettini et al. (\cite{pett00}) themselves argue that 
the presence of P Cygni features due to massive stars supports a history of 
continuous star formation, with an Initial Mass Function (IMF) which is 
Salpeter up to $M \sim 100\,{\rm M_\odot}$.  De~Mello et al. (\cite{deme00}) 
use the strengths of both O and B star features to argue that continuous star 
formation has proceeded for 25--100\,Myr, an age range whose lower end agrees 
with the SED fit of Ellingson et al. (\cite{elli96}).  For a metallicity $Z = 
0.4\,\mathrm{Z_\odot}$ (comparable to the $Z \simeq 0.32\,\mathrm{Z_\odot}$ 
measured by Teplitz et al. \cite{tepl00}), the appropriate \emph{Starburst99} 
models of Leitherer et al. (\cite{leit99}) predict that the intrinsic ratio 
$R_0 = F_{1475}/F_\mathrm{H\alpha}$ will rise from $\sim 98$ to $\sim 120$ 
over the course of the 25--100\,Myr interval.  \object{cB58} has 
an observed $R = 90$; the fact that $R < R_0$ can be attributed to the fact 
that extinction is higher in the UV than in the optical.  Using the Calzetti 
$k^\prime(\lambda)$, the LMC $k(\lambda)$, and the relation between color 
excesses, we can derive a single equation in $E_s(B-V)$:
\begin{eqnarray}
\mathrm{log}\,(R/R_0) & = & 0.4\,E_s(B-V) \times \\
                      &   & \Big(k(6563\,\mathrm{\AA})/0.44 - 
k^\prime(1475\,\mathrm{\AA})\Big) \nonumber
\end{eqnarray}
whose solution is $E_s(B-V) = 0.042 \pm 0.021$.  Such low values are 
\emph{required} to avoid discordant UV and $\mathrm{H\alpha}$-based estimates 
of the star formation rate in \object{cB58}.

\vspace{1.0mm}
\noindent
%\begin{underline}
{\bf IV. The UV spectral slope}
%\end{underline}
%\subsection{The UV spectral slope}

According to the empirical relation derived by Meurer et al. (\cite{meur99}), 
$A(1600\,\mathrm{\AA}) = 1.99(\beta - \beta_0)$ for $\beta_0 = -2.23$.  Using  
the Calzetti $k^\prime(\lambda)$, we then derive $E_s(B-V) = 0.190 \pm 0.028$.
From the analogous relation between $\beta$ and $E_s(B-V)$ in Calzetti 
(\cite{calz00}) and her suggestion of $\beta_0 = -2.3$, we would obtain 
$E_s(B-V) = 0.236 \pm 0.032$.

\section {Conclusions}

Although our calculations of $E_s(B-V)$ in \object{cB58} take uncertain 
measurements and apply scaling relations about which there is substantial 
local scatter, it remains striking that three of the four estimates are in 
excellent agreement with each other.  Only the UV spectral slope gives a value 
which is too high by a factor of 2--3.  In principle, use of a steeper 
(e.g., SMC) extinction curve rather than the Calzetti $k^\prime(\lambda)$ 
would lower the color excess estimated from the 
observed $\beta - \beta_0$.  However, a steeper extinction curve would also 
reduce the values of $E_s(B-V)$ derived from $F_\mathrm{FIR}/F_{1600}$ and 
population synthesis models; the difference is real.  

The preponderance of evidence for a low dust content in \object{cB58} 
argues against analyses which would describe its location in 
Figure~\ref{fig:meur} solely in terms of a far-IR deficit (e.g., Sawicki 
\cite{sawi01}; van~der~Werf et al. \cite{vand01}).  We can erase an apparent 
shortfall in $F_\mathrm{FIR}/F_{1600}$, for example, if we predict 
$F_\mathrm{FIR}$ from $\nu f_\nu(321.8\,\mathrm{\mu m})$ under the assumption 
that the dust in \object{cB58} is as hot as the dust in the most extreme of 
the local starbursts, \object{Tololo\,1924-416} ($T_\mathrm{d} = 
50.3\,\mathrm{K}$ for $\epsilon = 2$).  However, since none of the other 42 
systems has $T_\mathrm{d} > 40\,\mathrm{K}$, we would essentially be 
discarding the local starburst analogy for LBGs while failing to explain how 
two independent lines of evidence (the $\mathrm{H\alpha/H\beta}$ and 
UV/$\mathrm{H\alpha}$ ratios) could still favor a low $E_s(B-V)$.  Similarly, 
we might imagine that the observed far-IR-to-UV ratio is depressed from its 
intrinsic value because UV emission reddened by dust along a favored line of 
sight is more highly magnified than the bulk of the far-IR emission.  This 
scenario too seems implausible.  Since the best lensing model of Seitz et al. 
(\cite{seit98}) predicts that a large fraction ($\sim 0.54 - 0.67$) of the 
background source is magnified into the \object{cB58} arc, and since UV 
continuum and $\mathrm{H\alpha}$ emission have broadly similar spatial 
distributions in local starbursts (Conselice et al. \cite{cons00}), we would 
require exceptional tuning of the lensing geometry to produce agreement in all 
estimates of $E_s(B-V)$ \emph{except} that from $\beta$.

The most straightforward conclusion about \object{cB58} is that its offset 
from the local relation in Figure~\ref{fig:meur} has more to do with $\beta$ 
than with $F_\mathrm{FIR}/F_{1600}$-- i.e., that the local starburst analogy 
holds more robustly for all other properties of \object{cB58} than it does for 
the UV spectral slope.  Observational systematics may be partly responsible: 
a spectroscopic $\beta$ can be artificially reddened by differential 
refraction, while photometry can mislead due to intergalactic absorption-- 
as in the case, perhaps, of the $g$ band magnitude for \object{cB58} used by 
Ellingson et al. (\cite{elli96}) to derive $E_s(B-V) \sim 0.3$.  More 
seriously, $\beta$ may simply be unreliable as a precise indicator of dust 
content for individual objects.  One obvious concern is that different recipes 
for measuring $\beta$ yield different results for the same objects.  
\object{cB58} is a case in point, with estimates ranging from our $-1.28 \pm 
0.14$ to -0.8 (Pettini et al. \cite{pett00}) to $-0.74 \pm 0.1$ (van~der~Werf 
et al. \cite{vand01}).  However, inconsistencies also exist at $z = 0$; 
measurements of $\beta$ from \emph{IUE} spectra by Heckman et al. 
(\cite{heck98}) and Meurer et al. (\cite{meur99}), for example, differ by up 
to 0.77 for common sources.  We have at least minimized the vulnerability of 
our own analysis to systematic errors by measuring $\beta$ with exactly 
the same prescription used for the comparison sample in Figure~\ref{fig:meur}.
However, given that our understanding of $\beta$ in \object{cB58} seems no 
more (if not less) secure than our understanding of the source's far-IR SED, 
sweeping conclusions about the long-wavelength properties of individual LBGs 
which depend heavily on their UV spectral slopes appear premature.

Recent measurements of the extragalactic far-IR background (e.g., Hauser et 
al. \cite{haus98}; Fixsen et al. \cite{fixs98}; Lagache et al. \cite{laga99}) 
have placed upper bounds on the luminosity density due to star 
formation as a function of redshift.  These constraints have stimulated 
competing claims as to what fractions of the background are due to optically 
detectable star formation (e.g., Ouchi et al. \cite{ouch99}; Adelberger \& 
Steidel \cite{adel00}), optically invisible star formation (e.g., Barger et 
al. \cite{barg00}), and dust-enshrouded nuclear activity (e.g., Almaini et al. 
\cite{alma99}).  While most of the global properties of \object{cB58} do 
appear to obey the same scaling relations as local starbursts-- encouraging 
news for the scaled-up starburst hypothesis for LBGs-- the misleading redness 
of its UV spectral slope offers a cautionary example of how extinction 
corrections and extrapolations of SEDs to long wavelengths can go awry (see 
also Adelberger \& Steidel \cite{adel00}).  For the moment, the true 
contribution of UV-selected samples to the far-IR background must therefore 
remain uncertain until we can detect enough LBGs in the rest far-IR to 
determine how closely they resemble \object{cB58} and/or local starbursts as 
a group.

\begin{acknowledgements}

We thank Frank Bertoldi, Robert Zylka, and the staff of the IRAM 30\,m for 
help with the observations and data reduction.  We also acknowledge helpful 
interactions with Tim Heckman, Claus Leitherer, Gerhardt Meurer, Max Pettini, 
Stella Seitz, and Christy Tremonti.  This research has made use of the 
NASA/IPAC Extragalactic Database (NED), which is operated by the Jet 
Propulsion Laboratory, California Institute of Technology, under contract with 
the National Aeronautics and Space Administration.

\end{acknowledgements}

\end{document}